\documentclass[superscriptaddress,twocolumn,floatfix,aps,preprintnumbers,showpacs]{revtex4}

\usepackage{graphicx,color}
\usepackage{amsmath,amssymb}
\usepackage{natbib}
\usepackage{color}
\usepackage[normalem]{ulem}

\renewcommand{\aa}{\hat{a}}
\newcommand{\adagger}{\hat{a}^\dagger}

\newcommand{\ppsi}{\hat{\psi}}
\newcommand{\psidagger}{\hat{\psi}^\dagger}
\newcommand{\n}{\hat{n}}
\newcommand{\N}{\hat{N}}

\usepackage[caption=false]{subfig}
\usepackage{graphics,amssymb,amsmath,epsfig,color}

\newcommand{\eqnref}[1]{Eq.~(\ref{#1})}

\begin{document}

\title{Backaction-Driven Transport of Bloch Oscillating Atoms in Ring Cavities}

\author{J.~Goldwin}
\affiliation{Midlands Ultracold Atom Research Centre, School of Physics and Astronomy,
University of Birmingham, Edgbaston, Birmingham B15 2TT, UK}

\author{B.~Prasanna~Venkatesh}
\affiliation{
Department of Physics and Astronomy, McMaster University, 1280 Main Street West, Hamilton, Ontario, Canada L8S 4M1}
\affiliation{Asia Pacific Center for Theoretical Physics, San 31, Hyoja-dong, Nam-gu, Pohang, Gyeongbuk 790-784, Korea.}

\author{D.~H.~J.~O'Dell}
\affiliation{
Department of Physics and Astronomy, McMaster University, 1280 Main Street W., Hamilton, Ontario, Canada L8S 4M1}

\date{\today}

\begin{abstract}
We predict that an atomic Bose-Einstein condensate strongly coupled to an intracavity optical lattice can undergo resonant tunneling and directed transport when a constant and uniform bias force is applied. The bias force induces Bloch oscillations, causing amplitude and phase modulation of the lattice which resonantly modifies the site-to-site tunneling. For the right choice of parameters a net atomic current is generated. The transport velocity can be oriented oppositely to the bias force, with its amplitude and direction controlled by the detuning between the pump laser and the cavity. The transport can also be enhanced through imbalanced pumping of the two counter-propagating running wave cavity modes. Our results add to the cold atoms quantum simulation toolbox, with implications for quantum sensing and metrology.
\end{abstract}

\maketitle

Periodic potentials play a prominent role in condensed matter systems, and highlight some of the fundamental differences between classical and quantum dynamics: a quantum particle undergoes strong scattering when its de~Broglie wavelength satisfies the lattice Bragg condition, and can undergo tunneling through classically forbidden regions between sites. Furthermore, if a constant bias force $F$ is applied a quantum particle is not transported in the direction of the force but instead performs Bloch oscillations with no net displacement at a frequency $\omega_{B}=Fd/\hbar$, where $d$ is the lattice period \cite{note1}. Indeed, transport of electrons in lattices with an applied DC electric field only occurs as a result of dephasing processes such as scattering from lattice defects.

Cold atoms present an especially attractive platform for studies of lattice systems because all of the critical parameters governing the dynamics are tunable in real time. In particular, it is possible to control tunneling and transport by modulating the potential in time. Transport in statistical phase space has been demonstrated in pulsed lattices, realizing the quantum delta-kicked rotor \cite{Moo95} and leading to dynamical localization \cite{Moo94} and chaos-assisted tunneling \cite{Ste01}. Directed transport has been observed through ratchet effects in driven dissipative \cite{Sch03} and Hamiltonian lattices \cite{Sal09}. Tunneling control has been achieved through harmonic shaking of lattices without \cite{Eck09,Lig07,Wic12} and with \cite{Iva08,Sia08} a bias force. It is thus possible to control the superfluid--Mott insulator transition \cite{Eck05a,Zen09} and to induce macroscopic delocalization \cite{Alb09} and transport \cite{Hal10} of Bloch oscillating atoms. Recently photon-assisted tunneling \cite{Eck05b} has been studied in strongly correlated quantum gases \cite{Ma11,Che11}, and artificial vector gauge potentials have been generated \cite{Str12}.

What is missing in these schemes is backaction by the atoms upon the electromagnetic fields generating the lattice. Contrast this with the strong backaction effects seen in solids, such as lattice-phonon mediated Cooper pairing and the Meissner effect in superconductors. In principle an optically trapped atomic gas causes refraction of the lattice light, but extremely low densities render this negligible under normal conditions. An exception is inside a high finesse optical cavity, where the multi-pass effect can increase the effective optical path length by several orders of magnitude \cite{Ye03}, leading to a shift of the cavity resonance that depends on the density distribution of the atoms. If this shift is on the order of the cavity line width, the number of cavity photons is modified by the atomic wave function and {\it vice versa}. This backaction leads to richer dynamics than is otherwise possible \cite{Mas05,Nie10,Rit13}, such as collective atomic recoil lasing \cite{Kru03,Cub04,Sla07} and single-photon bistability \cite{Gup07,Rit07}. In this context the system can be considered an application of cavity optomechanics \cite{Mur08,Bre08,Sch11}, where collective excitations of the atoms play the role of material oscillators which are dispersively coupled to one or more cavity modes \cite{Kan10}.

It has been shown theoretically that such systems allow continuous, non-destructive measurements of atomic Bloch oscillations through detection of intensity and phase modulation of the transmitted light \cite{Ped09,Ven09}. Despite the backaction on the lattice effected by the atoms, Bloch's acceleration theorem remains valid, and this modulation of the lattice potential occurs predominantly at the expected Bloch oscillation frequency (i.e., calculated for an equivalent static tilted lattice) and its harmonics. It is therefore natural to ask whether this dynamical modulation of the lattice can drive the renormalization of atomic tunneling which is now familiar from experiments with free-space lattices. The central result of this Letter is to show that backaction-driven modulation of an intracavity lattice can lead to directed atomic current with tunable magnitude and direction.

\begin{figure}[h]
\centering
\includegraphics[height=3cm]{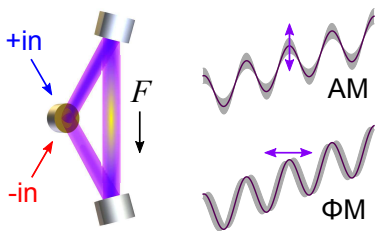}
\caption{An optical lattice is created by pumping the two running wave modes of a ring cavity. A trapped Bose-Einstein condensate (yellow ellipse) undergoes Bloch oscillations due to the bias force $F$. Atomic backaction leads to lattice amplitude and phase modulation (AM and $\Phi$M, respectively), which in turn induces coherent directed transport of the condensate.}
\label{fig:fig1}
\end{figure}

To see how this happens, we consider a Bose-Einstein condensate trapped along one leg of an optical ring cavity, as shown schematically in Fig.~\ref{fig:fig1}. Transverse degrees of freedom are assumed to be frozen out by external confinement, effectively reducing the dynamics to a single spatial dimension $z$. The two running-wave modes of the cavity are pumped through a lossless input-output coupling mirror by a laser with frequency $\omega_0=ck_r$, where $c$ is the speed of light in vacuum and $\hbar k_r$ is the recoil momentum. The light is detuned far enough from the atomic resonance that the excited state of the atoms can be adiabatically eliminated. For simplicity we also ignore atomic collisions, which may be negligible in an experiment either because the scattering cross-section is naturally small \cite{Alb09}, or has been made small through the use of a tunable Feshbach resonance \cite{Hal10}. In a frame rotating at $\omega_0$, and in the dipole and rotating wave approximations, the Hamiltonian is then given by
\begin{eqnarray} \nonumber
\hat{H} &=& -\hbar\!\!\sum_{k=\pm}\!\!\left[\Delta_c\,\adagger_k\aa_k + i(\eta_k^\ast\aa_k-\eta_k\adagger_k)\right] \\ 
&+&\!\! \int\!\!\mathrm{d}z\,\,\psidagger\!\!\left(-\frac{\hbar^2}{2m}\,\partial_z^2 + \hbar U_0\hat{\mathcal{E}}^\dagger\hat{\mathcal{E}} - Fz \right)\!\ppsi 
\label{eq:H}
\end{eqnarray}
where the annihilation operators $\aa_{+}$ and $\aa_{-}$ acting on the cavity modes, and $\ppsi(z)$ acting on the atomic field, all obey bosonic commutation relations. 
$\Delta_c=\omega_0-\omega_c$ is the detuning of the driving laser from the bare cavity resonance frequency $\omega_c$, and $\eta_k=\sqrt{J_k\kappa/2}$ for an incident flux of $J_k$ photons per unit time and a photon number decay rate of $2\kappa\langle\adagger_k\aa_k\rangle$ for each mode. The dimensionless positive-frequency component of the electric field is given by $\hat{\mathcal{E}}(z,t)=\aa_{+}\exp(ik_rz)+\aa_{-}\exp(-ik_rz)$. The depth of the lattice is proportional to $U_0$, which is a function of the atomic dipole moment, the cavity mode volume and the detuning from atomic resonance, and $F<0$ is the uniform and constant bias force.

We can alternatively express the cavity modes in a standing wave basis, described by the annihilation operators $\aa_c=(\aa_{+}+\aa_{-})/\sqrt{2}$ and $\aa_s=i(\aa_{+}-\aa_{-})/\sqrt{2}$~, where the $c$ ($s$) mode has cosine (sine) spatial symmetry. In this basis the interaction term in the second line of (\ref{eq:H}) takes the form
\begin{eqnarray}
\label{eq:Hi}
\hat{H}_i &=& \hbar U_0\left[\n\N + (\n_c-\n_s)\,\mathcal{C} + (\adagger_c\aa_s + \adagger_s\aa_c)\,\mathcal{S}\right]
\end{eqnarray}
where $\n=\n_c+\n_s=\adagger_c\aa_c+\adagger_s\aa_s$ is the total number of photons, $\N$ is the number of atoms in the condensate, and $\mathcal{C}[\ppsi]=\langle\cos(2k_rz)\rangle$ and $\mathcal{S}[\ppsi]=\langle\sin(2k_rz)\rangle$ depend implicitly on the atomic state. We will be most interested in cases where $\langle{\n_c}\rangle\gg\langle{\n_s}\rangle$, so that the intracavity lattice has predominantly cosine symmetry. Then the quantity $\mathcal{C}$ characterizes the degree of spatial ordering of the atoms and $\mathcal{S}$ is related to the coherence between the lowest and first excited Bloch bands \cite{Ped09}. Viewed in the optomechanical picture $\mathcal{C}$ and $\mathcal{S}$ represent the occupations of the lowest momentum modes of the condensate, with $\mathcal{S}=0$ in the absence of Bloch oscillations or a symmetry-breaking optical bistability \cite{Che10}. However even if we choose the $\eta_k$ to initially give $\mathcal{S}=0$, during Bloch oscillations $\mathcal{S}$ becomes nonzero and the intensity and spatial phase of the lattice vary dynamically.

To solve the full nonlinear dynamics we write the Heisenberg-Langevin equations, $i\hbar\partial_t\aa_\mu=[\aa_\mu,\hat{H}]-i\hbar\kappa\,\aa_\mu$ for $\mu=c,s$ and $i\hbar\partial_t\ppsi=[\ppsi,\hat{H}]$, ignoring all input noise operators, whose means are zero, and neglecting atom losses over the time scales of interest so that $N=\langle\hat{N}\rangle$ is constant. Letting $\alpha_\mu=\langle\aa_\mu\rangle$ and $\psi=\langle\ppsi\rangle/\sqrt{N}$, and factoring the expectation values of operator products, we obtain the mean-field equations,
\begin{eqnarray}
\label{eq:alpha1}
\partial_t\,\alpha_c &=& -(\kappa-i\Delta_+)\,\alpha_c - iU_0N\mathcal{S}\alpha_s + \eta_c \\
\label{eq:alpha2}
\partial_t\,\alpha_s &=& -(\kappa-i\Delta_-)\,\alpha_s - iU_0N\mathcal{S}\alpha_c + \eta_s \\
\label{eq:psi}
i\hbar\,\partial_t\,\psi &=& \left(-\frac{\hbar^2}{2m}\,\partial_z^2+\hbar U_0|\mathcal{E}(z)|^2 - Fz\right)\psi
\end{eqnarray}
Here $\Delta_\pm=(\Delta_c-U_0N)\mp U_0N\mathcal{C}$ are the effective detunings, and $\mathcal{E}(z,t)=\langle\hat{\mathcal{E}}(z,t)\rangle$ is the dimensionless electric field. The standing-wave modes are pumped at rates $\eta_c=(\eta_{+}+\eta_{-})/\sqrt{2}$ and $\eta_s=i(\eta_{+}-\eta_{-})/\sqrt{2}$~. Again we see that the lattice modulation is driven by changes in $\mathcal{C}$ and $\mathcal{S}$ during Bloch oscillations --- $\mathcal{C}$ drives amplitude modulation through changes of the Stark detuning $\Delta_+$ of the dominant cosine mode, and $\mathcal{S}$ induces shaking of the lattice (i.e., phase modulation) through coherent coupling of the sine and cosine modes. 

\begin{figure}[h]
\centering
\includegraphics[width=8.8cm]{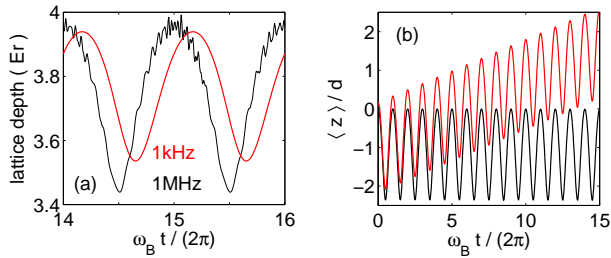}
\caption{Backaction induced lattice modulation and atomic transport. Only the cosine cavity mode was pumped, with $U_0=-2\pi\times 1$~Hz and $\Delta_c=U_0N=-\kappa$. (a) Lattice depth in units of the recoil energy $E_r=\hbar\omega_r$ as a function of time during intracavity Bloch oscillations. The black curve is for $\kappa=2\pi\times 1$~MHz, and the red curve for 1~kHz. (b) Condensate centroid position as a function of time, showing \emph{uphill} transport for $\kappa\sim\omega_B$. Colors are the same as in (a).}
\label{fig:fig2}
\end{figure}

In Fig.~\ref{fig:fig2}, we show the dynamics for $^{88}$Sr atoms accelerating under gravity in a 689~nm ring cavity lattice (i.e., lattice spacing $d=\pi/k_r=344.5$~nm). In this case we have $\omega_B = 2\pi\times 745$~Hz, and the recoil frequency $\omega_r\equiv\hbar k_r^2/(2m)=2\pi\times 4.78$~kHz. Here we choose $U_0N=-\kappa$ corresponding to the onset of collective strong coupling, and compare the dynamics with $\kappa=2\pi\times 1$~kHz and 1~MHz. Bloch oscillations induce lattice amplitude modulation of $\sim\!10\%$ peak-to-peak in both cases, with negligible shaking. The fast oscillations on top of the main modulation observed for larger $\kappa$ were identified previously, and are predominantly higher harmonics of $\omega_{B}$ associated with the nonlinearity of the coupled atom-light system \cite{Ven09}. In the case of small $\kappa$ these features are outside the cavity bandwidth and therefore suppressed.

Despite similar modulation depths for small and large $\kappa$ in Fig.~\ref{fig:fig2}, transport is only observed when $\kappa$ is on the order of $\omega_B$.  This is because $\kappa$ is the relaxation rate for the light field, which in turn sets the relative lag between the backaction-induced lattice modulation and the atomic Bloch oscillation. When $\kappa \gg \omega_{B}$ the light adapts almost instantaneously to the atomic motion at $\omega_{B}$ and there is consequently no delay between the two; when $\kappa \sim \omega_B$ the light is not able to adiabatically follow the atoms' motion and a phase lag develops, as evident in Fig.\  \ref{fig:fig2}(a). The effect of such a delay on transport in externally driven free-space lattices is well known \cite{Hal10,Kud11}, but in such cases the phase difference is a controlled parameter. In a cavity, on the other hand, the phase difference comes about self-consistently according to Eqs.~(\ref{eq:alpha1})--(\ref{eq:psi}).

Neglecting for the moment the origin of the modulation in Fig.~\ref{fig:fig2}, the resulting transport is similar to that in a free-space optical lattice under modulation of the lattice amplitude or phase, or of the bias force $F$ \cite{Tho02,Alb10,Tar12}. Physically, transport occurs because the phase lag ensures that the duration for which the lattice is shallower, and therefore tunneling more effective, overlaps more with the motion in one direction than in the other. By breaking this symmetry, the band center is effectively shifted from zero quasimomentum and a net motion occurs \cite{Sup}. The transport velocity is given by the group velocity $v_{g}=\hbar^{-1} \partial E(q) /\partial q$ averaged over a full Bloch oscillation, where $E(q)$ is the dispersion relation of the atoms in the untilted lattice, and $q$ is the quasimomentum \cite{Kud11}. If the phase lag is zero, $v_{g}$ averages to zero by the symmetry of $E(q)$ about $q=0$, but for a finite lag the lattice modulation falls out of sync with the atoms, leading to transport at velocity
\begin{eqnarray}\label{eq:vt}
v_t  &=& - d\,T_1\sin \phi
\end{eqnarray}
where $T_1$ is the tunneling rate between neighboring sites, which is proportional to the modulation depth, and $\phi$ is the lag between the modulation and the Bloch oscillation. For initial site-to-site coherence, such as we consider here, the result is directed transport superposed on the underlying Bloch oscillation \cite{Hal10}. Given initially random site-to-site phases, one instead observes spatial spreading of the atoms \cite{Iva08,Alb09}.

\begin{figure}[h]
\centering
\includegraphics[width=8.8cm]{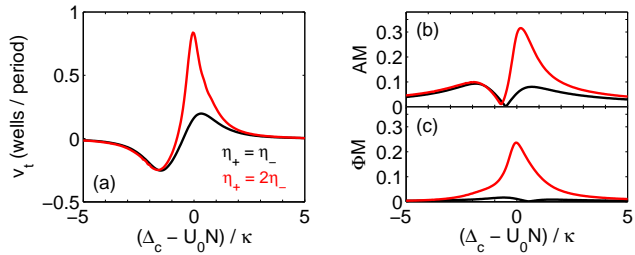}
\caption{Control of transport with detuning. (a) Transport velocity as a function of $\Delta_c$, for the same parameters as Fig.~\ref{fig:fig2} and $\kappa=2\pi\times 1$~kHz. The black curve is for balanced pumping, and the red curve for a strong imbalance. (b) Relative depth of self-induced lattice amplitude modulation (AM), defined as half the peak-to-peak lattice depth modulation, normalized to the mean lattice depth. (c) Phase modulation ($\Phi$M) defined as half the peak-to-peak spatial extent of the lattice shaking, in units of the lattice period.}
\label{fig:fig3}
\end{figure}

We have extended the above theory for free-space lattices to include the self-consistent optomechanical effects of a cavity. Under the approximation of nearest-neighbor tunnelling, which is valid for modulation at $\omega_B$, we find analytic expressions for  $T_1$ and $\phi$ which are given in the supplementary material \cite{Sup}. The analytic theory is in excellent agreement with the numerical simulations.
The magnitude and direction of transport depend on the cavity pumping parameters $\Delta_c$ and $\eta_{\pm}$, as well as the Stark shift $U_0N$. We find that $|v_t|$ increases quadratically with $U_0N$ for small values, and is maximized for intermediate lattice depths ($\sim 3\hbar\omega_r$ for the parameters of Fig.~\ref{fig:fig2}). In very shallow lattices, tunneling is strong but the modulation depth is reduced in proportion to the trap depth; in deep lattices tunneling is suppressed and the modulation depth is reduced due to flattening of the Bloch bands. The dependence on detuning $\Delta_c$ is shown in Fig.~\ref{fig:fig3}. For balanced pumping, with $\eta_+=\eta_-$ ($\eta_s=0$), the transport exhibits a dispersive shape around the Stark-shifted cavity resonance ($\Delta_+=0$). This is because modulation of $\mathcal{C}$ during the Bloch oscillation effectively dithers the detuning of the cosine cavity mode, as described by Eq.~(\ref{eq:alpha1}). The result is a modulation amplitude which approximately follows the derivative of the Lorentzian cavity response. The modulation phase also changes across the resonance, adding to the detailed shape we observe. Again these effects are well captured by the analytic tight-binding theory. We note in passing that for larger blue detunings, Raman transitions of the type studied in \cite{Wol12} can become resonant, leading to Rabi oscillations between condensate momentum states and tunneling into higher bands.

Because the effects we have described so far are dominated by lattice amplitude modulation, they are qualitatively present in standing wave cavities as well. However, only in ring cavities is it possible to pump the counterpropagating running-wave modes with independent amplitudes, corresponding to direct pumping of the sine mode (not to be confused with a translation of the lattice, such as occurs during lattice shaking). In Fig.~\ref{fig:fig3}(a) we observe a strong enhancement of uphill transport around $\Delta_c=U_0N$ when $\eta_{+}=2\eta_{-}$, even though the initial trap depth is the same as before. As seen in Figs.~\ref{fig:fig3}(b) and (c), both the amplitude and phase modulations are increased. Imbalanced pumping increases population of the cavity sine mode, thereby enhancing the backaction induced lattice shaking. At first it may be surprising that this effect is only pronounced around $\Delta_c=U_0N$, but the effective detuning of the cosine mode $\Delta_+$ becomes positive here, corresponding to the regime of cavity heating \cite{Gan00}, where linear stability analysis for $F=0$ predicts unstable dynamics \cite{Hor01}. In contrast, near the positive transport peak one finds that both the sine and cosine modes operate in the cavity cooling regime with $\Delta_\pm$ negative, where the $F=0$ dynamics are damped.

\begin{figure}[h]
\centering
\includegraphics[width=8.8cm]{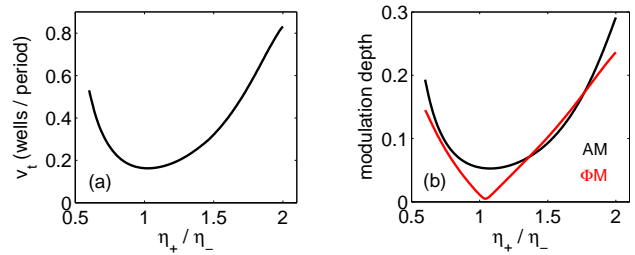}
\caption{Effects of imbalanced pumping of the running-wave cavity modes. (a) Transport velocity as a function of $\eta_{+}/\eta_{-}$, with $\Delta_c=U_0N$ and the same parameters as Fig.~\ref{fig:fig3}. (b) Amplitude and phase modulation depths as defined above.}
\label{fig:fig4}
\end{figure}

The effect of imbalanced pumping is investigated further in Fig.~\ref{fig:fig4}, where we vary the ratio $\eta_{+}/\eta_{-}$, for $\Delta_c=U_0N$ and fixed initial trap depth. Phase modulation is more sensitive than amplitude modulation to small pump asymmetries. We observe that the transport minima are slightly offset from the condition of balanced pumping; the actual minima occur where the time-averaged value of $\mathcal{S}$ over a full Bloch oscillation period vanishes. This is due to weak pumping of the sine mode balancing the atomic dynamics. Simulations at strong imbalances reveal the existence of numerous types of instability. These include Landau-Zener tunneling, symmetry-breaking bistability \cite{Che10}, and collective excitations of the condensate \cite{Hor01}, and will be the subject of a future work \cite{VenXX}. Near the onset of instability, the depths of amplitude and phase modulation as defined here become comparable; recall Figs.~\ref{fig:fig3}(b) and (c). However we note that AM is more efficient at driving transport when each type of applied modulation is considered in isolation for free-space lattices. Transport appears to be dominated by amplitude modulation in all of the parameter regimes we have studied, with a four-fold increase in $|v_t|$ possible through imbalanced pumping of the cavity. 

We can now compare our results to experiments in driven free-space lattices.  Combined Bloch oscillations and transport have been observed by imaging atoms both \emph{in situ} and in time-of-flight \cite{Alb09,Hal10}. For intracavity lattices one could detect the Bloch oscillations non-destructively in the transmitted light \cite{Ped09,Ven09}. In fact, transport has its own unique signature: for the case shown in Fig.\ \ref{fig:fig2} we find a factor of two imbalance between the spectral power of the cavity fields in the sidebands at $\pm \omega_{B}$. Indeed, transport up/down in our system is analogous to sideband-resolved cavity optomechanical heating/cooling \cite{vuletic00,kippenberg08}. For a pump detuning $\Delta_+\sim+\omega_B$, the $-\omega_B$ sideband is near resonance with the cavity and therefore dominates over the further detuned $+\omega_B$ sideband; the atoms undergo energy conserving Raman transitions up the Wannier-Stark ladder \cite{Tar12} of states separated in energy by steps of $\hbar \omega_{B}$. Conversely, a red-detuned pump with $\Delta_+\sim-\omega_B$ leads to a dominant $+\omega_B$ sideband, with the atoms descending the Wannier-Stark ladder.

An important feature of the backaction driven dynamics is that $\omega_{B}$ appears to remain unchanged from its value in a static lattice. Because the lattice in a ring cavity can accelerate, this does not have to be the case \cite{Sam14}. That the system oscillates at $Fd/\hbar$ is critical for applications in metrology and sensing where the modulation frequency is taken as a measure of the force \cite{Ped09}. It also implies that backaction will not induce so-called super Bloch oscillations \cite{Hal10}, because these occur when the lattice modulation is detuned from $\omega_{B}$. It is worth noting that although the backaction driven $\Phi$M is much smaller than what has been applied in free-space experiments \cite{FM}, the values of AM we observe are similar to what was applied in Refs.~\cite{Alb10,Tar12}. Finally, we note that our tight-binding theory predicts that in the absence of initial site-to-site coherence, the backaction induced modulation will decay. This is because $\mathcal{C}$ and $\mathcal{S}$ become constant, cutting off the modulation of the lattice according to Eq.~(\ref{eq:Hi}).

In conclusion, we have shown that optomechanical effects lead to qualitatively new dynamics for a Bose-Einstein condensate undergoing Bloch oscillations in a high finesse optical cavity. As the condensate quasimomentum samples the first Brillouin zone, the optical lattice depth and position are dynamically modulated, even as the Bloch frequency itself is unchanged. When the cavity damping rate is on the order of the Bloch oscillation frequency, coherent directed transport of the condensate can be observed. Asymmetric pumping of the running wave cavity modes enhances the transport, and in extreme cases makes the system dynamically unstable. Our results extend the study of coherent control of tunneling to include optomechanical lattice excitations. They are relevant to measurements of Bloch oscillations in cavities and other non-destructive atomic probes, and also more generally to attempts to realize neutral atom quantum simulators where backaction upon electromagnetic fields is significant.

\emph{Note added}: since the preparation of this manuscript, we have become aware of the experimental observation of transport of a Bloch oscillating Bose-Einstein condensate in a high-finesse standing wave cavity \cite{HemPC}.

\begin{acknowledgments}
We acknowledge stimulating discussions with P. Courteille and A. Hemmerich, and are grateful to the anonymous referees for useful suggestions. This work was funded by NSERC in Canada and EPSRC in the United Kingdom, the Max Planck Society, the Korea Ministry of Education, Science and Technology, Gyeongsangbuk-Do, Pohang City, and the National Research Foundation of Korea Basic Science Research Program No. 2012R1A1A2008028.
\end{acknowledgments}


%
%

\pagebreak
\widetext
\clearpage 
~\vspace{2cm} 
\begin{center}
\textbf{\large Supplemental Material}
\end{center}
\setcounter{equation}{0}
\setcounter{figure}{0}
\setcounter{table}{0}
\setcounter{page}{1}
\makeatletter
\renewcommand{\theequation}{S\arabic{equation}}
\renewcommand{\thefigure}{S\arabic{figure}}
\renewcommand{\bibnumfmt}[1]{[S#1]}
\renewcommand{\citenumfont}[1]{S#1} 

\begin{center}
\parbox{14cm}{We provide an analytical treatment of the coupled atom-light equations for Bloch oscillations in a ring cavity and compare the predictions of the theory to numerical simulations.}
\end{center}
~\vspace{1cm}

In this supplement we provide an analytical theory describing the driven transport of Bloch oscillating atoms in a ring cavity. Our approach is to extend and adapt the theory developed in \cite{Sthommenph,Sthommenanalyt} for transport in externally driven optical lattices to the situation where the drive arises from the coupled atom-light dynamics.

Before describing the full coupled atom-cavity dynamics, we briefly show how transport can arise in the case of atoms undergoing Bloch oscillations in an amplitude-modulated lattice without backaction. To do so we consider the lowest Bloch band, characterized in the standard tight-binding model by a dispersion relation $E(q)=-J\cos(\pi q)$, where $q$ is the qusimomentum, and $2J$ is the band width, which only depends on the lattice depth. The transport velocity is given by the group velocity averaged over a full Bloch period \cite{SKudo11}, giving
\begin{eqnarray}
v_t &=& \frac{1}{\hbar k_r}\,\left.\frac{\partial E(q)}{\partial q}\right|_{q_0}
\end{eqnarray}
where $\hbar k_r$ is the recoil momentum as in the main text, and $q_0=0$ is the initial quasimomentum.

\begin{figure}
\includegraphics[width=0.45\textwidth]{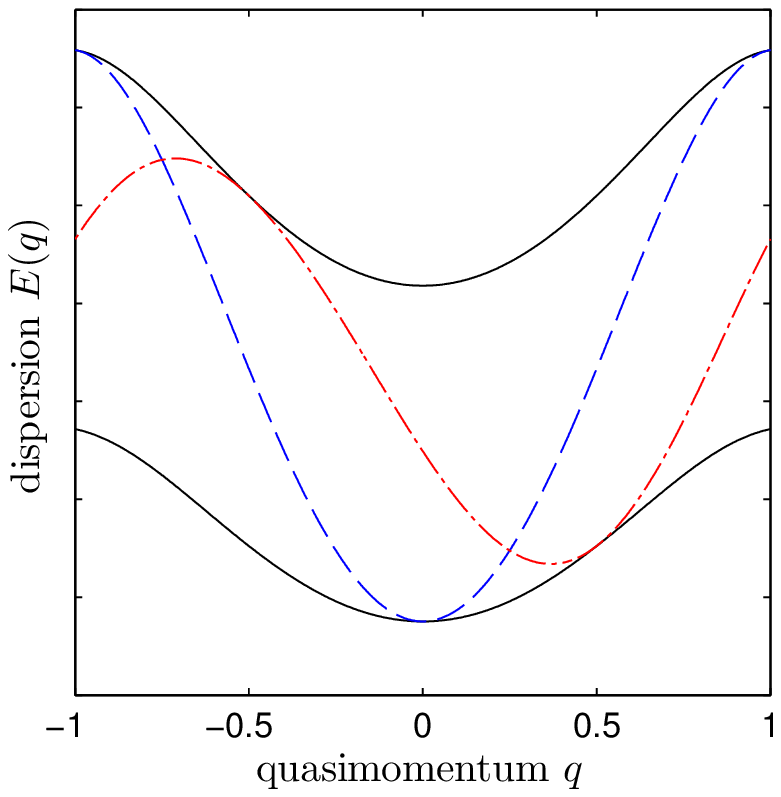}
\caption{Transport through band modulation. The upper and lower solid black curves show bands for lattices with fixed depth $-3$ and $-4$, respectively, in units of the recoil energy. The blue dashed curve shows the dispersion relation for a lattice which is modulated in-phase with the Bloch oscillation ($\phi=0$), and the red dash-dottted curve is for out-of-phase lattice modulation ($\pi=\pi/2$). The transport velocity is proportional to the slope at zero quasimomentum.}
\label{fig:band}
\end{figure}

Now suppose the lattice depth $V$ is modulated at the Bloch oscillation frequency, so that $V=V(q)$. Then the bandwidth $J$ is also modulated, so that
\begin{eqnarray}
\hbar k_r v_t &=& -\cos(\pi q_0)\,\left.\frac{\partial J(q)}{\partial q}\right|_{q_0} + \pi J(q_0)\sin(\pi q_0) \\
&=& -\left.\frac{\partial J}{\partial q}\right|_{q_0}
\end{eqnarray}
For a modulation of the form $V=V_0[1+\epsilon\cos(\pi q + \phi)]$, a simple Taylor expansion of $J(V)$ around $V_0$ reveals that
\begin{eqnarray}
v_t &\approx& \frac{-\epsilon V_0 J^\prime\pi}{\hbar k_r}\sin(\phi)
\label{eq:vt_toy}
\end{eqnarray}
where $J^\prime=\partial J/\partial V|_{V_0}$. Thus the transport velocity is proportional to the modulation depth and can be positive, negative, or zero, depending on the phase. This is shown schematically in Fig.~\ref{fig:band}. In-phase modulation ($\phi=0$) preserves the symmetry of the modulated band, so that the transport remains zero. In contrast, out-of-phase modulation ($\phi=\pi/2$) shifts the center of the band so that at $q=0$ the slope of the dispersion, and therefore the transport velocity, is no longer zero.

The above discussion has assumed an applied lattice modulation. We now describe what happens when the modulation arises through the coupled atom-cavity dynamics, so that the modulation depth and phase are not controlled externally. To that end we write the equations of motion for the atomic meanfield in a scaled form (energies scaled by $E_r = \hbar \omega_r$, times scaled by $\omega_r^{-1}$ and lengths scaled by $1/k_r$):
\begin{align}
i \partial_t \psi &= H(t) \psi = \left(-\partial_z^2 + s(t) \cos^2\left[z-\frac{\theta(t)}{2}\right] - f z \right) \psi \label{eq:scaledatom}\\
s(t) &= 2 U_0 \sqrt{(n_c-n_s)^2+4 \Re(\alpha_c^{*}\alpha_s)^2}, \,\, \tan\left[\theta(t)\right] = \frac{2 \Re(\alpha_c^*\alpha_s)}{n_c - n_s}. \label{eq:latdepphase} 
\end{align}
In the above equation we have retained the same notation as the main paper for the scaled single atom dispersive shift $U_0$ and the position coordinate $z$. The scaled force is given by $f = \omega_B/\left(\pi \omega_r\right)$. The trap depth $s(t)$ and spatial phase shift $\theta(t)$ are extracted from  $U_0 \vert \mathcal{E} (z,t) \vert^2$ after ignoring $z-$independent terms. \eqnref{eq:scaledatom} clearly shows that the dynamics of the light field leads to a phase and amplitude modulated intracavity optical lattice. In this note we restrict ourselves to the situation where the phase dynamics is insignificant and the transport mainly arises from the trap depth modulation $s(t)$. This assumption is valid for much of the parameter regime we have studied. From the numerical solutions presented in Fig.~1 of the paper and from the analytical considerations presented here we find the trap depth as a function of time has the following form:
\begin{align}
s(t) = s_i e^{-2 \kappa t} + s_0 (1-e^{-2\kappa t}) + \Delta s \cos (\omega_B t + \phi) \label{eq:latdepfnt}.
\end{align}
The transient part of the above equation comes from the damping of the chosen initial trap depth $s_i$. For the Bloch periodic part we have chosen to ignore the higher harmonics since we find from our numerical simulations they provide relatively small contributions in comparison to the fundamental. The steady state mean trap depth $s_0$, the modulation amplitude $\Delta s$, and the phase shift $\phi$, are determined self-consistently from the coupled atom-field dynamics. In what follows we will compute analytical expressions for these parameters which together with the findings of \cite{Sthommenph,Sthommenanalyt} determines the resulting transport. 

In order to tackle the field equations (Eqs.~(3) and (4) in the paper) we need two inputs from the atomic dynamics: $\mathcal{C} = \langle \cos(2z) \rangle$ and $\mathcal{S} = \langle \sin(2z) \rangle$. Following \cite{Sthommenph,Sthommenanalyt} we will calculate these variables by expanding the atomic state in the Wannier-Stark (WS) basis. Let us first split the hamiltonian (\ref{eq:scaledatom}) in the following form and define the WS states $\varphi_n(z)$ (restricted to the lowest band):
\begin{align}
H(t) &= H_0 + H_1(t) \nonumber\\
H_0 \varphi_n (z) &= \left(-\partial_z^2 + \frac{s_0}{2} \cos\left(2z\right) - f z \right) \varphi_n (z) = e_n \varphi_n(z) \label{eq:WSdefn}\\
H_1(t) &= \frac{\Delta s}{2}\cos (\omega_B t + \phi) \cos(2z), 
\end{align}
where the time-dependent part of the hamiltonian is understood to be valid for times $t \gg \kappa^{-1}$ when transients have damped away. In any case the exact form of the trap depth modulation is irrelevant for the calculation of $\mathcal{C},\mathcal{S}$ within the nearest neighbor approximation to be used below. The WS states $\varphi_n(z)$ are localized around the $n$th cell and the energies form a ladder separated by the Bloch frequency with $e_m-e_n = (m-n)\omega_B$. Expanding the atomic mean field in the WS basis we have:
\begin{align}
\psi(z,t) &= \sum_{n} d_n(t) e^{i \phi_n(t)} \varphi_n(z)\\
\phi_n(t) &= n \omega_B t + \gamma_0 \int_0^t d \tau \frac{\Delta s}{2}\cos (\omega_B \tau + \phi)
\end{align}
where the WS overlaps are defined as:
\begin{align}
\gamma_p = \gamma_{-p} &= \langle \varphi_n \vert \cos(2z)\vert \varphi_{n+p} \rangle= \langle \varphi_{n+p} \vert \cos(2z)\vert \varphi_{n}\rangle\\
\nu_p = \nu_{-p} &= \langle \varphi_n \vert \sin(2z)\vert \varphi_{n+p} \rangle = \langle \varphi_{n+p} \vert \sin(2z)\vert \varphi_{n} \rangle.
\end{align} 

The equations of motion for the coefficients $d_n$ are given by:
\begin{align*}
i \dot{d_n} = \frac{\Delta s}{2}\cos (\omega_B t + \phi) \sum_{p \neq 0} d_{n+p} e^{i p \omega_B t} \gamma_p. 
\end{align*}
The above equation can be solved by going into a Fourier representation: $d(k,t) = \sum_{n=-\infty}^{\infty} d_n(t) e^{ink}$ \cite{Sthommenanalyt}. Making use of the form of the time-dependent wave function we can write for the required atomic inputs $\mathcal{C,S}$:
\begin{align}
\langle \cos(2z) \rangle &= \gamma_0 + 2 \gamma_1 \vert \sigma_1 \vert \cos(\omega_B t + \theta_1) \label{eq:cos2x}\\
\langle \sin(2z) \rangle &= \nu_0 + 2 \nu_1 \vert \sigma_1 \vert \cos(\omega_B t + \theta_1)\label{eq:sin2x}\\
\sigma_1 &= \vert \sigma_1 \vert e^{i\theta_1} = \sum_n d_n^*(t) d_{n+1}(t) \label{eq:coherence}.
\end{align}
In the equation above we have made our first crucial approximation: due to the spatial localization of WS states, $\gamma_p$ and $\nu_p$ go to zero quickly as $p$ increases and in the {\it nearest neighbor} approximation we restrict ourselves to just $p=\pm 1$. As a result the overlap integrals depend only on the site-to-site coherence in the WS basis, $\sigma_1$ defined in \eqnref{eq:coherence}. This can easily be evaluated using the Fourier space solution as:
\begin{align}
\sum_{n} d_n^* d_{n+1}(t) = \frac{1}{2\pi} \int_{-\pi}^{\pi} dk e^{ik} \vert d(k,t) \vert^2 = \frac{1}{2\pi} \int_{-\pi}^{\pi} dk e^{ik} \vert d(k,0) \vert^2 = \sum_{n} d_{n}^* d_{n+1} (t=0) \label{eq:coherenceexp}
\end{align}
where we have used the property that the time evolution leaves the magnitude of $d(k,t)$ constant \cite{Sthommenanalyt}. Thus $\{\mathcal{C,S}\}$ are dependent only on the initial site to site coherence of the wave function in the WS basis. Since the specific Bloch periodic form of $\Delta s(t)$ is not crucial to derive the above result for $\{\mathcal{C,S}\}$, it can also include the transient part of the trap depth evolution. In this case the initial coherence is something we control by the choice of the initial wave function. Given a WS basis we can compute $\sigma_1$ easily. Before we proceed to tackle the light field evolution a few remarks are in order. Using the above solution in the WS basis, we can evaluate the position expectation for long times in the nearest neighbor approximation ($t \gg \kappa^{-1}$) as (\cite{Sthommenanalyt,Sthommenamp}):
\begin{align}
\langle z (t) \rangle - \langle z  (t = 0) \rangle =  \vert \sigma_1 \vert z_1 \cos(\omega_B t +\theta_1) + T_1  t \sin(\theta_1-\phi) \label{eq:wavepktposn}
\end{align}
with $z_1 = \langle \varphi_n \vert z \vert \varphi_{n+1} \rangle$ and $T_1 = \frac{\Delta s}{2} \vert \sigma_1 \vert \gamma_1$. The above equation clearly shows that on top of the Bloch periodic evolution there is a constant velocity drift of magnitude (in units of lattice periods per unit time):
\begin{align}
v_t =  T_1  \sin(\theta_1-\phi) \label{eq:transptvel}.
\end{align}  

Another thing to make note of is the fact that the overlaps, $\{\nu_0,\nu_1 \}$, of the function $\sin(2z)$  are smaller than the respective overlaps for the cosine function, $\{\gamma_0,\gamma_1\}$. The reason for this is that $\sin(2z)$ is anti-symmetric about each well's center and edge whereas the WS wave function's density and adjacent site overlaps are approximately even \footnote{This can be understood by expanding the WS wave functions in terms of Wannier functions. In the limit of a large enough force, the WS states are well localized and the largest contribution comes from a single Wannier function which has a definite parity about the center of the well.}. Thus $\mathcal{S}$ is small throughout the evolution for symmetric pumping ($\eta_+ = \eta_-$) and provided we start with $\alpha_s = 0$. In addition since we stay at reasonably small values of $\vert N U_0 \vert \sim O(\kappa)$, we can completely ignore the role of $\alpha_s$ in the field dynamics. In this limit the trap depth is simply given by $s(t) \approx 2 U_0 n_c(t)$. Let us now look at the field dynamics for $\alpha_c(t)$ while ignoring the sine mode:
\begin{align}
\partial_t \alpha_c = -(\kappa - i \Delta_+)\alpha_c + \eta_c.\label{eq:fieldeqn}
\end{align}
Using \eqnref{eq:cos2x}, we have $\Delta_+ = \delta - u_0-u_1 \cos(\omega_B t+ \theta_1)$, where $\delta = \Delta_c-NU_0$, $u_0 = NU_0 \gamma_0$ and $u_1 = 2 NU_0\gamma_1 \vert\sigma_1\vert$. For $t \gg \kappa^{-1}$ we can solve \eqnref{eq:fieldeqn} to give:
\begin{align}
\alpha_c(t) &= \eta_c e^{-i \frac{u_1}{\omega_B}\sin(\omega_Bt+\theta_1)} \int_{-t}^0 d \tau e^{\kappa \tau - i (\delta-u_0)\tau} \exp\left(i \frac{u_1}{\omega_B} \sin \left[\omega_B t+\omega_B \tau + \theta_1 \right]\right).\label{eq:fieldeqnsoln} 
\end{align}

We can expand the exponential inside the integral in terms of Bessel functions $J_n(y)$ using the Jacobi-Anger expansion $e^{iy\sin(\beta)} = \displaystyle \sum_{n=-\infty}^{\infty} J_n(y)e^{in\beta}$ \footnote{When comparing with numerical simulations on finite but large lattice sizes (we typically use about 64 lattice sites), the sums here have to be understood as approximate. But since typically the argument of the Bessel function in the chosen parameter regimes are small, we can extend the sum to infinity.} and compute the final result as:
\begin{align}
\alpha_c(t) = \eta_c e^{-i \frac{u_1}{\omega_B}\sin(\omega_Bt+\theta_1)} \displaystyle \sum_n J_n\left(\frac{u_1}{\omega_B}\right)\frac{e^{in(\omega_B t+\theta_1)}}{\kappa - i \delta_n} \label{eq:fieldsoln},
\end{align}
with $\delta_n = \delta_0 - n\omega_B = \delta - u_0 - n\omega_B$. Although the driving from $\Delta_+$ in our approach has only the fundamental Bloch frequency, the atom-field non-linear interaction gives rise to harmonics of the Bloch frequency in the field response above. Since the higher harmonics are relatively small, we will neglect them. From the above solution we can write down the oscillating trap depth for $t \gg \kappa^{-1}$:
\begin{align}
\frac{s(t)}{2U_0} &=  \eta_c^2 \sum_n \frac{J_n\left(\frac{u_1}{\omega_B}\right)^2}{\kappa^2 + \delta_n^2} +  \eta_c^2 \left( X_s^{*}e^{-i(\omega_Bt+\theta_1)} + X_s e^{i(\omega_Bt+\theta_1) }\right) \label{eq:latdepsoln}\\
X_s &= \sum_n \frac{J_n\left(\frac{u_1}{\omega_B}\right)J_{n-1}\left(\frac{u_1}{\omega_B}\right)}{(\kappa-i\delta_n)(\kappa+i\delta_{n-1})} = \vert X_s \vert e^{-i\mu_s}\label{eq:latdeposc}
\end{align}
Comparing the above expression with \eqnref{eq:latdepfnt} we immediately have for the mean lattice depth:
\begin{align}
s_0 = 2 U_0 \eta_c^2 \sum_n \frac{J_n\left(\frac{u_1}{\omega_B}\right)^2}{\kappa^2 + \delta_n^2}. \label{eq:s0fixpoint}
\end{align}
We can determine $s_0$ by solving the above equation as a fixed point equation. It is then easy to evaluate and relate the oscillatory parts of \eqnref{eq:latdepfnt} and \eqnref{eq:latdeposc}: $\Delta s = 4 U_0 \eta_c^2 \vert X_s \vert$ and $\phi = \theta_1 - \mu_s$. In this manner, given a set of cavity parameters and an initial atomic meanfield, we can compute the trap depth and atomic dynamics as functions of time.

\begin{figure*}
\centering
  \subfloat[Phase lag as a function of detuning]{\label{fig:phaselagfndc}
\includegraphics[width=.45\textwidth]{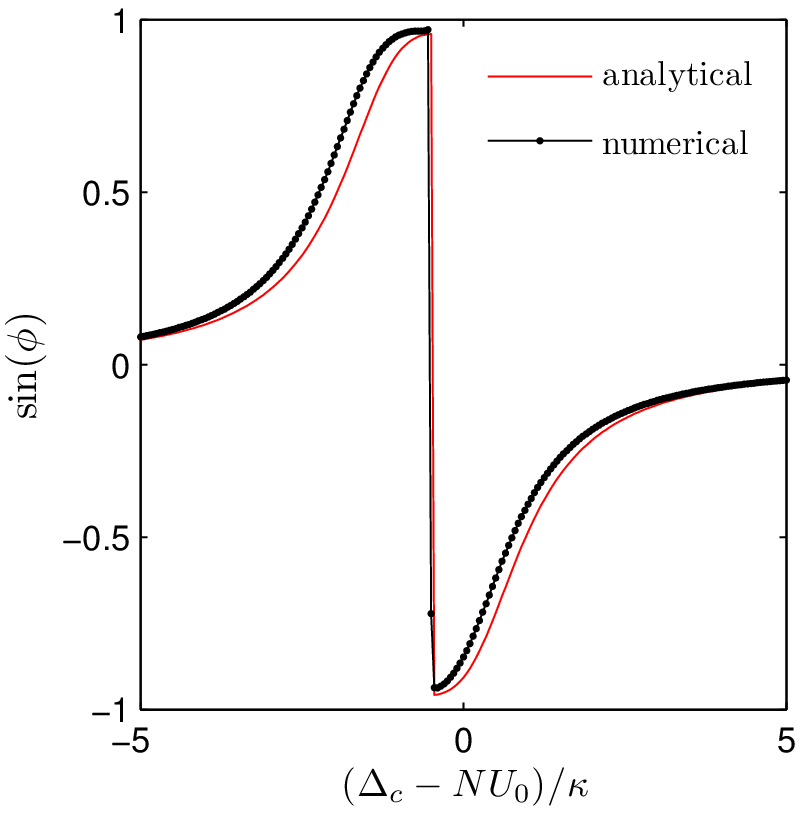}}
  \hfill 
  \subfloat[Phase lag as a function of cavity decay]{\label{fig:phaselagfnkap}\includegraphics[width=.45\textwidth]{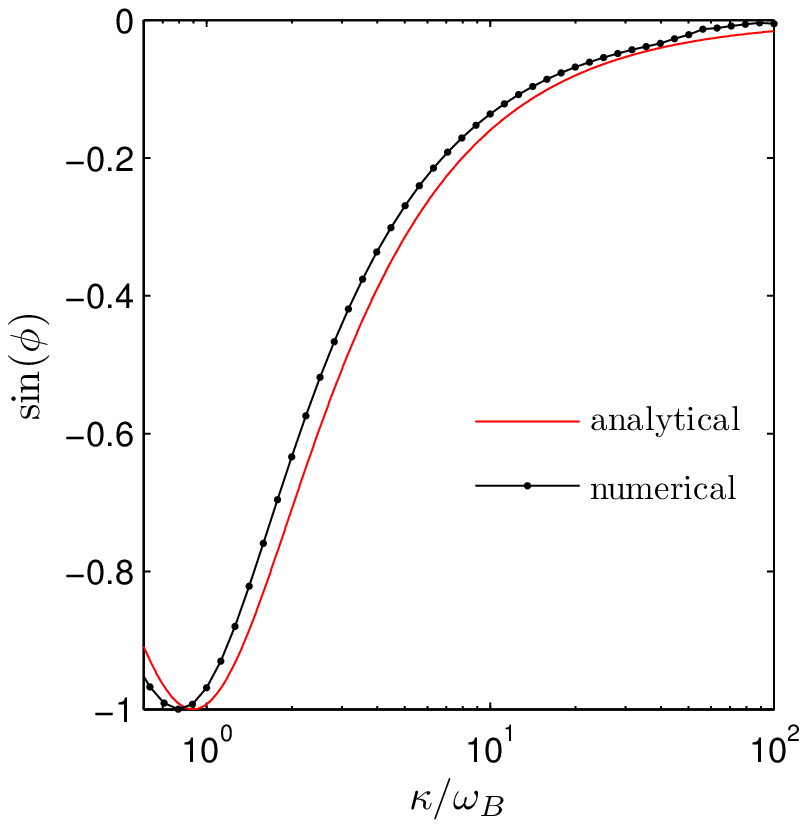}}
\caption{Phase lag between the lattice amplitude modulation and atomic Bloch oscillation dynamics as a function of detuning (a) and cavity decay (b). The red lines were computed using Eq.\,(\ref{eq:phaselagfndckap}) and the black dotted lines were extracted from numerical simulations. Parameters are same as in Fig.~3(a) of the main paper. In (a) the phase lag changes sign at $\delta_0 = 0$, indicating that transport velocity can be controlled by varying the detuning $\Delta_c$. In (b) as $\kappa$ is varied, $NU_0$ is scaled to maintain $NU_0/\kappa = -1$ and $\Delta_c = N U_0$. The phase lag is optimal for $\kappa \sim \omega_B$.}
\label{fig:phaselag}
\end{figure*}

Let us now compare this theory with the numerical results. We first consider the parameter regime where the cavity field evolution is much faster than the Bloch frequency i.e. $\omega_B \ll \{\kappa,N\vert U_0\vert,\Delta_c\}$. In this case we can approximate $\delta_n \approx \delta_{n \pm 1}$ in \eqnref{eq:latdeposc}, which immediately implies that $X_s$ is real and $\mu_s = 0$. This gives $\theta_1 = \phi$, leading to vanishing transport velocity \eqnref{eq:transptvel}. Thus, in contrast to the free space case \cite{Sthommenph}, irrespective of the choice of the initial coherence (which sets $\theta_1$) for $\omega_B \ll \kappa$ there is no transport in agreement with results of the numerical simulation shown in Fig.~2 of the main paper (see also Fig.~\ref{fig:phaselagfnkap} in this supplement). On the other hand when $\omega_B \gg \{\kappa, N\vert U_0 \vert,\Delta_c \}$, the argument of the Bessel functions in \eqnref{eq:latdeposc}, $u_1/\omega_B \ll 1$. This leads to vanishingly small trap depth modulation killing off transport. In the regime $\omega_B \sim \kappa$ these two extreme effects are absent and transport behavior can be seen as in the paper.

\begin{figure*}
\centering
  \subfloat[Position and lattice depth]{\label{fig:meanxcomp}
\includegraphics[width=.45\textwidth]{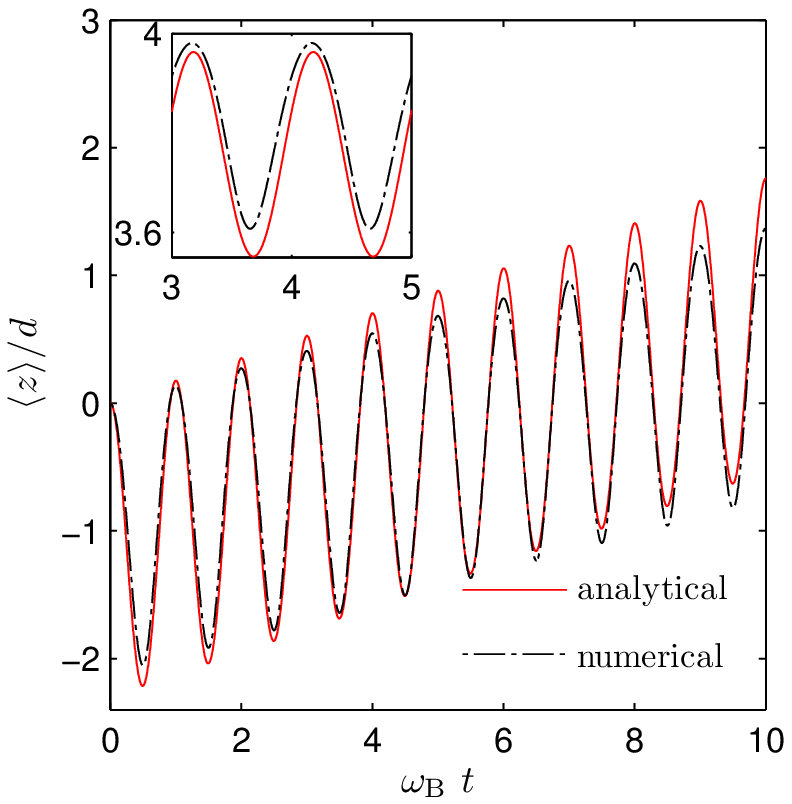}}
  \hfill
  \subfloat[Transport velocity as a function of detuning]{\label{fig:vtfndccomp}\includegraphics[width=.45\textwidth]{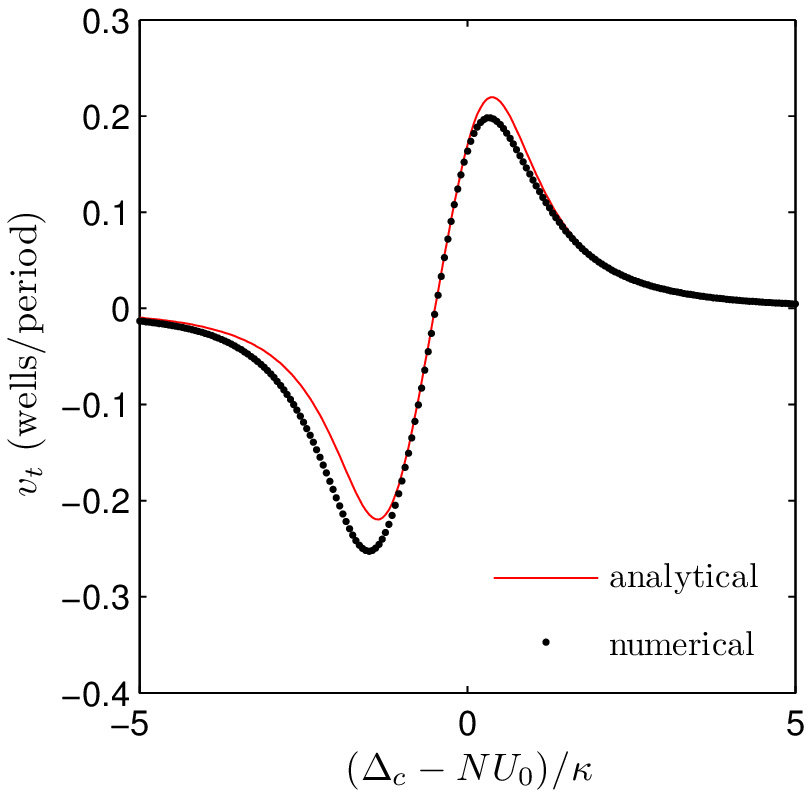}}
\caption{In (a), mean displacement and lattice depth (inset) obtained from the numerical simulation (solid red line) is compared to the analytical theory developed in this supplement (black dash dotted line). Other parameters are the same as in Fig.~2 of the main paper. For the chosen parameters in (a), the cavity detuning shifted by the atomic dynamics, $\Delta_c - N U_0 \langle \cos(2z)\rangle$, is positive resulting uphill transport (see text for further details). In (b) the transport velocity from the numerical simulation (black dots), shown in Fig.~3a of the paper, is compared to the analytical theory (red solid line) given by \eqnref{eq:transptvelfinal}. This demonstrates the validity of the analytical method developed here.}
\label{fig:numvsanalyt}
\end{figure*}

For the parameter regimes in the paper where $\omega_B \sim \kappa$, we still have $u_1/\omega_B < 1$ as $\vert \gamma_1\vert \sim O(10^{-2})$ due to the localized nature of WS states. Thus we can approximate the sum over Bessel functions in \eqnref{eq:latdeposc} by taking just $n=0,\pm 1$. This allows us to get the following result for the trap depth modulation amplitude and phase (with real initial wave function, giving $\theta_1 = 0$):
\begin{align}
\Delta s &= \frac{4 U_0 \eta_c^2 J_0(\frac{u_1}{\omega_B})J_1(\frac{u_1}{\omega_B})}{\left(\kappa^2+\delta_0^2 \right) \sqrt{\left(\kappa^2+\delta_1^2\right)\left(\kappa^2+\delta_{-1}^2\right)}} 2 \vert  \delta_0 \vert \omega_B \label{eq:latdepfndckap}\\
\sin(\phi) &= \frac{-2\kappa \omega_B \mathrm{\,\,sign}(\delta_0)}{\sqrt{\left(\kappa^2+\delta_1^2\right)\left(\kappa^2+\delta_{-1}^2\right)}} \label{eq:phaselagfndckap}
\end{align}
From the form of the phase lag $\phi$ above and \eqnref{eq:transptvel}, it is clear than we can control the sign of the transport velocity by the sign of $\delta_0$. In Fig.\,\ref{fig:phaselag} we compare the above analytical result for the phase lag as a function of detuning $\Delta_c$ and cavity damping $\kappa$ with numerical simulations. For small $u_1/\omega_B$, we can also truncate the sum in \eqnref{eq:s0fixpoint} and approximate $s_0 \approx 2 U_0 \eta_c^2 \frac{J_0\left(\frac{u_1}{\omega_B}\right)^2}{\kappa^2 + \delta_0^2}$, leading to the following expression for the transport velocity: 
\begin{align}
v_t = s_0 \gamma_1 \vert \sigma_1 \vert \frac{J_1\left(\frac{u_1}{\omega_B}\right)}{J_0\left(\frac{u_1}{\omega_B}\right)}\frac{4\omega_B^2 \kappa \delta_0}{(\kappa^2+\delta_1^2)(\kappa^2+\delta_{-1}^2)} \label{eq:transptvelfinal}
\end{align}
At a given value of $\Delta_c$, the above expression shows that although tunnelling increases with decreasing $s_0$, this will not always result in larger transport as noted in the paper. When the mean lattice depth $s_0$ is fixed, \eqnref{eq:transptvelfinal} can be used to understand the behaviour of transport velocity as a function of the detuning $\Delta_c$ shown in Fig. 3a. in the paper. Clearly the derivative of a Lorentzian-like shape of the transport velocity follows from the form of \eqnref{eq:transptvelfinal}. In Fig.~\ref{fig:meanxcomp} we compare atomic dynamics calculated from the analytical method developed here with numerical simulations presented in Fig.~2 of the main paper. The system parameters chosen are $\kappa = 1 $KHz, $NU_0/\kappa = -1$, $\Delta_c = NU_0$ and the parameters that go into the analytical calculation are $\{\gamma_0 = 0.50,\gamma_1 =-0.04, \sigma_1 = 0.98, \theta_1 = 0 \}$. The average cavity detuning shifted by atomic dynamics $\delta_0 = 0.5 \kappa$ is positive leading to the observed uphill transport in Fig.~\ref{fig:meanxcomp}. In \ref{fig:vtfndccomp} we compare the transport velocity as a function of detuning calculated from numerical solutions with the analytical calculation \eqnref{eq:transptvelfinal} and find good agreement.

Finally, some comments are in order regarding the situation with imbalanced pumping ($\eta_+ \neq \eta_-$). When the sine mode $\alpha_s$ is also pumped by the external driving laser, the potential extrema are shifted from $z = {n \pi,n \pi/2}$. This means that the WS functions will also be shifted which immediately leads to significant values of $\mathcal{S}$. Thus we can no longer ignore the sine mode dynamics. As a result it is not straightforward to extend our analytical scheme to this situation.

\begin{figure}
\centering
\includegraphics{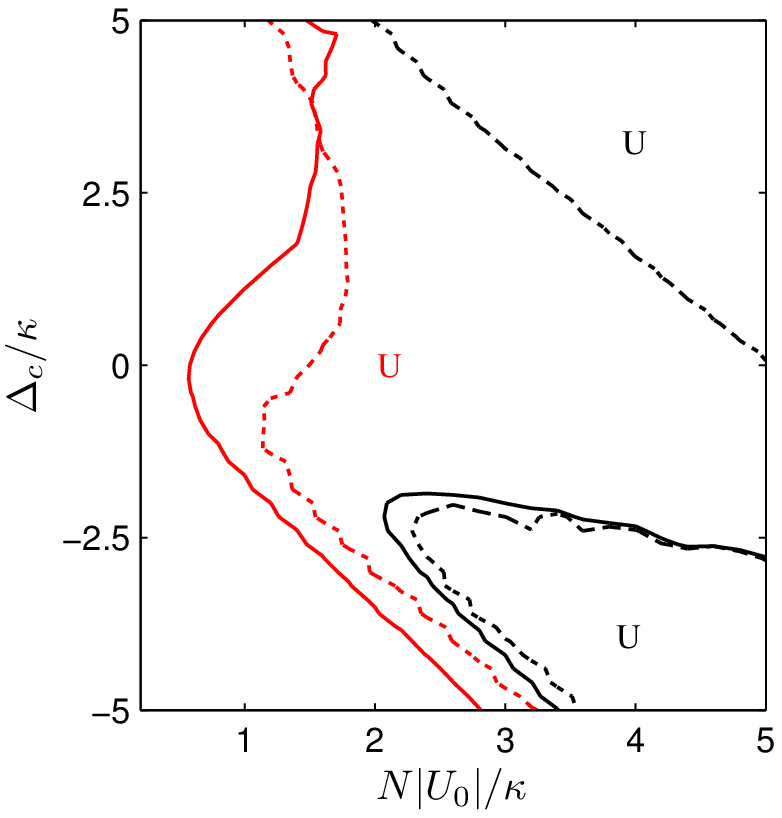}
\caption{Stability diagram. Black curves are for balanced pumping ($\eta_+=\eta_-$) and red are imbalanced ($\eta_+=2\eta_-$), with the initial trap depth fixed to $4E_r$ and all other parameters the same as in Fig.~2 of the main text (with $\kappa=2\pi\times 1$~kHz). Unstable regions are indicated by a `U' to the right of the boundaries. The solid and dash-dotted boundaries were determined by requiring that $|\alpha_c \pm i\alpha_s|$ and the Debye-Waller factor $\vert\langle e^{i2z} \rangle\vert^2$, respectively, become smaller than some threshold value (see text for details).}
\label{fig:stabplot}
\end{figure}

Moreover, we find that large regions of the parameter space exhibit instabilities for imbalanced pumping. From our numerical simulations, we identify the onset of unstable dynamics {\it via} two characteristic features. First, the Debye-Waller factor $\vert \langle e^{i2z} \rangle \vert^2$ quantifies the commensurability of the atomic density with the lattice period, and becomes very small when higher momentum states become populated. Second, when $\alpha_c(t) \sim \pm i \alpha_s$, one can see from Eq.~(\ref{eq:latdepfnt}) that the lattice depth becomes very small in magnitude, leading to a break down of stable Bloch oscillations. In Fig.~\ref{fig:stabplot} we have presented a stability diagram over detuning and coupling strength, computed from our numerical simulations according to these two stability criteria. For balanced pumping (black curves), we observe two unstable regions. Both regions show excitation of the condensate in the form of reduced Debye-Waller factors, which is a general feature of the instabilities we have observed. The lower unstable region also shows excitation of the sine cavity mode, which we associate with the symmetry-breaking bistability described in \cite{SChen10}. For imbalanced pumping (red curves) these two regions merge, significantly reducing the parameter space where stable dynamics occur. In future work we will extend the above analytical method to cases with strong imbalanced pumping to understand the many types of stable and unstable dynamics we have observed from numerical simulations \cite{Sbpvfuture}.

\end{document}